\documentclass[twocolumn,superscriptaddress,floatfix,preprintnumbers, nofootinbib,hyperref]{revtex4-2} 
\pdfoutput=1
\usepackage[normalem]{ulem}
\usepackage[utf8]{inputenc}
\usepackage{amsmath,amssymb}
\usepackage{epsfig}  
\usepackage{graphicx}   
\usepackage{slashed}       
\usepackage{tikz}
\usepackage{amsmath}
\usepackage{tikz-feynman}
\usepackage{url}
\usepackage{color}
\usepackage{xcolor}
\usepackage{multirow}
\usepackage{comment}
\usepackage{amssymb}
\usepackage{orcidlink}
\usepackage[version=3]{mhchem}
\usepackage{bm}
\usepackage{gensymb}
\DeclareUnicodeCharacter{2212}{-}
\DeclareMathOperator{\cm}{cm}
\DeclareMathOperator{\Hz}{Hz}
\DeclareMathOperator{\GeV}{GeV}

\DeclareMathOperator{\MeV}{MeV}

\DeclareMathOperator{\kpc}{kpc}

\allowdisplaybreaks

\begin{document}
\hypersetup{allcolors=[rgb]{0.0 0.0 0.6}, linkcolor=[rgb]{0.75 0.05 0.05}}

\title{Constraining gravitational-wave backgrounds from conversions into photons in the Galactic magnetic field}

\author{Alessandro~Lella~\orcidlink{0000-0002-3266-3154}}
\email{alessandro.lella@ba.infn.it}
\affiliation{Dipartimento Interateneo di Fisica  ``Michelangelo Merlin'', Via Amendola 173, 70126 Bari, Italy}
\affiliation{Istituto Nazionale di Fisica Nucleare - Sezione di Bari, Via Orabona 4, 70126 Bari, Italy}%

\author{Francesca~Calore~\orcidlink{0000-0001-7722-6145}}
\email{calore@lapth.cnrs.fr}
\affiliation{LAPTh, CNRS, F-74000 Annecy, France}
\affiliation{LAPP, CNRS, F-74000 Annecy, France}

\author{Pierluca~Carenza~\orcidlink{0000-0002-8410-0345}}\email{pierluca.carenza@fysik.su.se}
\affiliation{The Oskar Klein Centre, Department of Physics, Stockholm University, Stockholm 106 91, Sweden}

\author{Alessandro~Mirizzi~\orcidlink{0000-0002-5382-3786}}
\email{alessandro.mirizzi@ba.infn.it}
\affiliation{Dipartimento Interateneo di Fisica  ``Michelangelo Merlin'', Via Amendola 173, 70126 Bari, Italy}
\affiliation{Istituto Nazionale di Fisica Nucleare - Sezione di Bari, Via Orabona 4, 70126 Bari, Italy}%

\date{\today}
\smallskip

\begin{abstract}

High-frequency gravitational waves ($f \gtrsim  1$ MHz) may provide a unique signature for the existence of exotic physics. The lack of current and future gravitational-wave experiments sensitive at those frequencies leads to the need of  employing different indirect techniques. Notably, one of the most promising one is constituted  by graviton-photon conversions in magnetic fields. In this work, we focus on conversions of a gravitational-wave background into photons inside the Milky-Way magnetic field, taking into account the state-of-the-art models for both regular and turbulent components. We discuss how graviton-to-photon conversions may lead to imprints in the cosmic photon background  spectrum in the range of frequencies $f\sim10^{9}-10^{26}\,\Hz$, where the observed photon flux is widely explained by astrophysics emission models.  Hence, the absence of any significant evidence for a diffuse photon flux induced by graviton-photon conversions allows us to set stringent constraints on the gravitational-wave strain $h_c$, strengthening current astrophysical bounds by $\sim1-2$ orders of magnitude in the whole range of frequencies considered.
 
\end{abstract}

\maketitle

\section{Introduction}

The detection of a Gravitational Wave Background (GWB) at high frequencies $f\gtrsim$~MHz has been proposed as a smoking gun for exotic physics, since no astrophysical source is expected to produce Gravitational Waves (GWs) in such a frequency range~\cite{Rosado:2011kv,Sesana:2016ljz,Lamberts:2019nyk,Robson:2018ifk,KAGRA:2021kbb,KAGRA:2021vkt,Babak:2023lro}. Conversely, GWBs generated in the early Universe could be characterized by signals falling in this high-frequency range, allowing one to probe inflation~\cite{Grishchuk:1974ny,Starobinsky:1979ty,Rubakov:1982df,Bartolo:2016ami}, first-order phase transitions~\cite{Caprini:2015zlo,Caprini:2019egz,Hindmarsh:2020hop,Gouttenoire:2022gwi,Athron:2023xlk}, topological defects~\cite{Vilenkin:2000jqa, Blanco-Pillado:2017oxo, Auclair:2019wcv, Gouttenoire:2019kij,Servant:2023tua} and primordial black holes~\cite{Anantua:2008am,Dolgov:2011cq,Dong:2015yjs,Franciolini:2022htd,Gehrman:2022imk,Gehrman:2023esa} (see also Ref.~\cite{Aggarwal:2020olq} for a review).

However, probing GWs above the MHz, is way beyond the sensitivities of current GW experiments, such as LIGO, Virgo and KAGRA~\cite{LIGOScientific:2014qfs,LIGOScientific:2019vic}, and also out of the reach for future ground-based ones, as Einstein Telescope (ET)~\cite{Hild:2010id, Punturo:2010zz} and Cosmic Explorer (CE)~\cite{LIGOScientific:2016wof}. Therefore, different indirect-detection techniques have to be developed to probe high-frequency GWs. In this regard, an interesting possibility discussed in literature, consists in exploiting the so-called inverse Gertsenshtein effect, i.e.~the conversion of GWs into photons inside magnetic fields~\cite{Gertsenshtein:1962kfm,Macedo:1983wcr,Raffelt:1987im,Cruise:2012zz,Dolgov:2012be,Ejlli:2018hke,Ejlli:2019bqj,Ejlli:2020fpt,Cembranos:2023ere}. This phenomenon is completely analogous to axion-photon conversions~\cite{Raffelt:1987im}, widely used to probe the QCD axion and other axion-like particles. Hence, some pioneering works have recently analyzed the physics potential of experiments originally thought for axion searches to probe the parameter space of GWs in the high-frequency range~\cite{Domcke:2022rgu, Domcke:2023bat, Gatti:2024mde}.
Analogously, another intriguing opportunity deals with the detection of observable signatures related to GW conversions in cosmic magnetic fields, such as planetary magnetospheres~\cite{Liu:2023mll,Ito:2023nkq}, the Milky-Way magnetic field~\cite{Ramazanov:2023nxz, Ito:2023nkq}, extragalactic, primordial magnetic fields~\cite{Domcke:2020yzq} and magnetic fields in Galaxy clusters~\cite{He:2023xoh}.

In this work, we discuss the possibility of probing high frequency GWs in a vast frequency range, $10^{9}-10^{26}$~Hz, through GW-photon conversion inside the Galactic magnetic field. The diffuse photon flux originated by GW conversions can then be compared with observations of the Cosmic Photon Background (CPB) from radio to gamma-rays, in order to constrain the relic density of GWs. 
This topic has been recently studied in Ref.~\cite{Ito:2023nkq}, by comparing the expected flux of photons originated by GW conversions in the turbulent Galactic magnetic field to observations of the various diffuse photon backgrounds. In the present work, we build on these results and improve the treatment of the magnetic field adding the contribution of the regular component, which we describe by employing realistic models that fit synchrotron measurements~\cite{Jansson:2012pc}. Moreover, we consider astrophysical backgrounds contributing to the diffuse emission and perform an accurate fitting procedure to constrain the possible GWB converting into photons. This procedure allows us to improve current constraints on the strain associated to a diffuse GWB, strengthening by $\sim1-2$ orders of magnitude the results present in the previous literature in the entire frequency range considered.

This work is structured as follows. In Sec.~\ref{sec:Effect} we review the theory associated to the Gertsenshtein effect, considering separately the simplified treatment for a constant magnetic field. In Sec.~\ref{sec:MilkyWayConversions} we study graviton-photon conversions in the benchmark Milky-Way magnetic field model adopted in this work, discussing also the relative importance of the turbulent component. In Sec.~\ref{sec:Distortion} we present the measurements of the CPB in the 
range of frequencies considered as well as the state-of-the-art emission models used to fit the data. The constraints obtained by employing these observations are discussed in Sec.~\ref{subsec:Results}. Finally, in Sec.~\ref{sec:Discussion} we summarize our results and conclude.

\section{Graviton-photon conversions}
\label{sec:Effect}

The dynamics of the graviton and photon fields in a magnetized medium is generically described by the action~\cite{Ejlli:2018hke}
\begin{equation}
    S=\int d^4x\sqrt{-g}\,\left(\mathcal{L}_{\rm gr}+\mathcal{L}_{\rm em}\right)\,,
\label{Eq:Action}
\end{equation}
where we are assuming natural units $\hbar=c=1$ and a flat space-time metric signature $\eta={\rm diag}(1,-1,-1,-1)$. In this equation we have introduced the metric tensor determinant $\sqrt{-g}$ associated to a curved space-time metric $g_{\mu\nu}(x)$. In particular, $g_{\mu\nu}(x)$ can be expanded as
\begin{equation}
    g_{\mu\nu}(x)=\eta_{\mu\nu}+\frac{2}{M_{\rm P}}h_{\mu\nu}(x)\,
\end{equation}
in which $h_{\mu\nu}(x)$ encodes small gravitational perturbations around the flat space time metric $\eta_{\mu\nu}$, i.e.~${|h_{\mu\nu}|\ll1}$.
Given these definitions, the free-graviton
Lagrangian reads
\begin{equation}
    \mathcal{L}_{\rm gr}=\frac{M_{\rm P}^2}{2}R
\end{equation}
where $R$ is the Ricci tensor and $M_{\rm P}\simeq2.44\times10^{18}\,\GeV$ is the reduced Planck mass. On the other hand, the dynamics of the electromagnetic field $A_\mu(x)$ interacting with the graviton field is described by 
\begin{equation}
    \begin{split}
        \mathcal{L}_{\rm em}&=-\frac{1}{4}g^{\mu\alpha}g^{\nu\beta}F_{\mu\nu}F_{\alpha\rho}\\
        &+\int d^4x'A_{\mu}(x)\Pi^{\mu\nu}(x,x')A_{\nu}(x')\,,
    \end{split}
\end{equation}
where $F_{\mu\nu}=\partial_{\mu}A_\nu+\partial_{\nu}A_\mu$ is the electromagnetic field strength and $\Pi^{\mu\nu}(x,x')$ is the photon field polarization tensor describing photon propagation in a magnetized medium.

In the Coulomb gauge, a photon with energy $\omega$ is described by
\begin{equation}
    A_i(\mathbf{x},t)=\sum_{\lambda=x,y}\,A_{\lambda}(\mathbf{x})\epsilon^i_{\lambda}\,e^{-i\omega t}\,,
\end{equation}
where the sum runs over the two independent photon polarization states in the plane transverse with respect to the direction of motion, which we identify as the $z$-axis. Analogously, the graviton field in the transverse-traceless (TT) gauge can be expanded as
\begin{equation}
    h_{ij}(\mathbf{x}, t)=\sum_{\lambda=\times,+}\,h_{\lambda}(\mathbf{x})e^{ij}_{\lambda}\,e^{-i\omega t}\,,
\end{equation}
in which the spin-2 polarization tensors $\{e^{ij}_{\times},e^{ij}_{+}\}$ are chosen such that
\begin{equation}            
    e^{ij}_{\times}=\epsilon^i_{x}\epsilon^j_{y}+\epsilon^i_{y}\epsilon^j_{x}\,\,\,\,\,\,\,\,\,\,\,\,\,\,\,\,
    e^{ij}_{+}=\epsilon^i_{x}\epsilon^j_{x}-\epsilon^i_{y}\epsilon^j_{y}\,.
\end{equation}

Starting from the action in Eq.~\eqref{Eq:Action}, it is possible to derive the equations of motion (EoMs) of the full system~\cite{Gertsenshtein:1962kfm, Raffelt:1987im,Ejlli:2018hke}
\begin{equation}
    \left(i\frac{d}{dz}-\omega\right)\begin{pmatrix}
        h_+\\
        h_\times\\
        A_x\\
        A_y\\
    \end{pmatrix}
    =
    H\,\begin{pmatrix}
        h_+\\
        h_\times\\
        A_x\\
        A_y\\
    \end{pmatrix}\,,
\label{Eq:EOM}
\end{equation}
where $\omega$ is the energy of the beam and a one-dimensional propagation along the $z$-axis is considered. In this regard, we highlight that Eq.~\ref{Eq:EOM} derives from a one-dimensional reduction of Maxwell's equations describing electrodynamics in presence of an external graviton source term. As discussed in Ref.~\cite{Leroy:2019ghm,Millar:2021gzs,McDonald:2024uuh} for the axion case, a complete three-dimensional treatment could introduce non-negligible corrections to the one-dimensional approximation adopted in this work. In the one-dimensional case, the evolution of the four-component field is governed by the Hamiltonian
\begin{equation}
    H=\begin{pmatrix}
        0   &\mathcal{H}_{g\gamma}\\
        \mathcal{H}_{g\gamma} &\mathcal{H}_{\gamma\gamma}\\
    \end{pmatrix} \,\ ,
\end{equation}
where $\mathcal{H}_{\gamma\gamma}$ is a $2\times2$ matrix encoding the photon dispersion relation
\begin{equation}
    \mathcal{H}_{\gamma\gamma}=\begin{pmatrix}
        \Delta_{x}c_\phi^2+\Delta_{y}s_\phi^2   &[\Delta_{y}-\Delta_{x}]c_\phi s_\phi\\
        [\Delta_{y}-\Delta_{x}]c_\phi s_\phi    &\Delta_{y}c_\phi^2+\Delta_{x}s_\phi^2
    \end{pmatrix}\,.
\end{equation}
In this expression $s_\phi$ and $c_\phi$ refer to the sine and the cosine of the angle between the transverse component of the magnetic field $\mathbf{B_T}$ and the orientation of the $\bm{\epsilon}_x$ 
polarization vector, i.e. $c_\phi=\mathbf{B_T} \cdot \bm{\epsilon_x}/|\mathbf{B_T}||\bm{\epsilon_x}|$ and $s_\phi=\sqrt{1-c_\phi^2}$.
The term $\Delta_{x}$ ($\Delta_{y}$) takes into account medium effects on the dispersion relation of the $x$ ($y$) photon polarization state 
\begin{equation}
    \Delta_{\lambda}=\Delta_{\rm pl}+\Delta^{\lambda}_{\rm QED}+\Delta_{\rm CMB}\,,
\end{equation}
where $\lambda=x,y$. The different terms are related to the effective photon mass in a plasma, QED vacuum effects and scattering on Cosmic Microwave Background (CMB) photons, respectively. Explicitly, they read~\cite{Latorre:1994cv,Kong:1998ic,Dobrynina:2014qba}
\begin{equation}
    \begin{split}
        \Delta_{\rm pl}=&-\frac{\omega_{\rm pl}^2}{2\,\omega}\\
        \simeq&\,-1.1\times10^{-3}\,\left(\frac{\omega}{1\,\MeV}\right)^{-1}\left(\frac{n_e}{10^{-2}\,\cm^{-3}}\right)\,\kpc^{-1}\,,\\
        \\
        \Delta^{\lambda}_{\rm QED}=&\,\kappa_\lambda\,\frac{4\,\alpha^2 \,B_T^2\,\omega}{45\,m_e^4}\\
        \simeq&\,4.5\times10^{-12}\,\kappa_\lambda\,\left(\frac{\omega}{1\,\MeV}\right)\left(\frac{B_T}{1\mu{\rm G}}\right)^2\,\kpc^{-1}\,,\\
        \\
        \Delta_{\rm CMB}=&\,\frac{44\pi^2\,\alpha^2\,T_{\rm CMB}^4\,\omega}{2025\,m_e^4}\\
        \simeq&\,8.7\times10^{-11}\,\left(\frac{\omega}{1\,\MeV}\right)\,\kpc^{-1}\,,
    \end{split}
\end{equation}
where $\kappa_\lambda=7/2,\,2$ for $x$ and $y$ photon polarization states, respectively, $\omega_{\rm pl}=\sqrt{4\pi\alpha n_e/m_e}$ is the plasma frequency, $n_e$ and $m_e$ are the electron number density and the free electron mass, and $T_{\rm CMB}\simeq2.73\,$K~\cite{Fixsen:2009ug} is the CMB temperature.
Notice that, in general,  $\Delta^{x,y}_{\rm QED}\ll\Delta_{\rm CMB}$ and $\Delta_x\simeq\Delta_y$. Thus, in the following we will drop polarization indices. The off-diagonal term
\begin{equation}
    \mathcal{H}_{g\gamma}=\begin{pmatrix}
        \Delta_{g\gamma} s_\phi   &\Delta_{g\gamma} c_\phi\\
        \Delta_{g\gamma} c_\phi   &-\Delta_{g\gamma} s_\phi\\
    \end{pmatrix}\,,
\end{equation}
causes graviton-photon mixing. 
In this expression we introduced
\begin{equation}
    \Delta_{g\gamma}=\frac{B_{\rm T}}{\sqrt{2}\,M_{\rm P}}\simeq8.8\times10^{-10}\left(\frac{B_{\rm T}}{1\,\mu{\rm G}}\right)\,\kpc^{-1}\,.
\end{equation}
We highlight that Eq.~\eqref{Eq:EOM}
holds as long as mixing particles are relativistic and their dispersion relation is given by $k=\sqrt{\omega^2-\omega_{\rm pl}^2}\simeq\omega$, where $k$ is the particle wave number~\cite{Raffelt:1987im,Battye:2019aco}. In this work we are interested in the range of frequencies $f\in[10^9,10^{26}]\,\Hz$, corresponding to photon energies $4\,\mu{\rm eV}\lesssim\omega\lesssim4\,\GeV$. On the other hand, for typical Milky-Way electron densities ${n_e\sim10^{-2}\,\cm^{-3}}$, plasma frequencies are always in the range $\omega_{\rm pl}\sim10^{-3}-10^{-2}\,$neV. Thus, the approximation leading to the
EoMs considered is always valid in our framework. Moreover, Eq.~\eqref{Eq:EOM} holds for photon energies $\omega\lesssim100\,$TeV, where absorption effects due to scattering over CMB and Extra-Galactic Background Light~(EBL) photons is negligible~(see Refs.~\cite{Mirizzi:2009aj,Dobrynina:2014qba,Kartavtsev:2016doq} for some works on this topic).\\
Since, in general, $\left[H(z),H(z')\right]\neq0$ for $z\neq z'$, Eq.~\eqref{Eq:EOM} does not admit analytical solutions. Therefore, the system has to be solved numerically, considering the evolution from an initial state made of unpolarized gravitons $(1/\sqrt{2},1/\sqrt{2},0,0)^{T}$ at $z\simeq20\,\kpc$. Then, the probability to have graviton-to-photon conversions along the trajectory of the beam towards the observation point at $z=0$ is given by
\begin{equation}
    P_{g\gamma}=|A_x(0)|^2+|A_y(0)|^2\,.
\end{equation}

\subsection{Homogeneous magnetic fields}
\label{subsec:HomogeneousField}
If the magnetic field is assumed to be constant and homogeneous along the line of sight, then the evolution equations significantly simplify. In this case, assuming a propagation along the $z$ axis, the $x$ mode of the photon field can be taken to be parallel to $\mathbf{B_T}$. In this case, it is possible to separate the $4\times4$ differential equation system into two independent $2\times2$ systems that can be solved analytically. In particular, it can be shown that $\{A_{x},h_{\times}\}$ and $\{A_{y},h_{+}\}$ polarization states couple separately. In particular, 

\begin{equation}
    \left[i\frac{d}{dz} -\omega\right]\begin{pmatrix}
                                            A_{x}\\
                                            h_{\times}
                                        \end{pmatrix}
            =\begin{pmatrix}
            &\Delta   & \Delta_{g\gamma}\\   
            &\Delta_{g\gamma}                      & 0\,
        \end{pmatrix}   \begin{pmatrix}
                            A_{x}\\
                             h_{\times}
                            \end{pmatrix}  \,,
    \label{Eq:EOMhom}
\end{equation}
and the same holds for $\{A_{y},h_{+}\}$. In this case, propagation equations can be solved analytically by diagonalizing the mixing matrix.
Therefore, the conversion probability simply reads~\cite{Raffelt:1987im}
\begin{equation}
    \begin{split}
        P_{g\gamma}&=\left|\langle A_\lambda|h_\lambda\rangle \right|^2\\
        &=\frac{4\Delta_{g\gamma}^2}{\Delta_{\rm osc}^2}\,\sin^2{\left(\frac{\Delta_{\rm osc}\,z}{2}\right)}\,,
    \end{split}
\label{Eq:ProbConstB}
\end{equation}
where 
\begin{equation}
    \Delta_{\rm osc} =\sqrt{\Delta^2+4\Delta_{g\gamma}^2}\,
\end{equation}

\begin{figure}[t!]
    \centering
    \includegraphics[width=1\columnwidth]{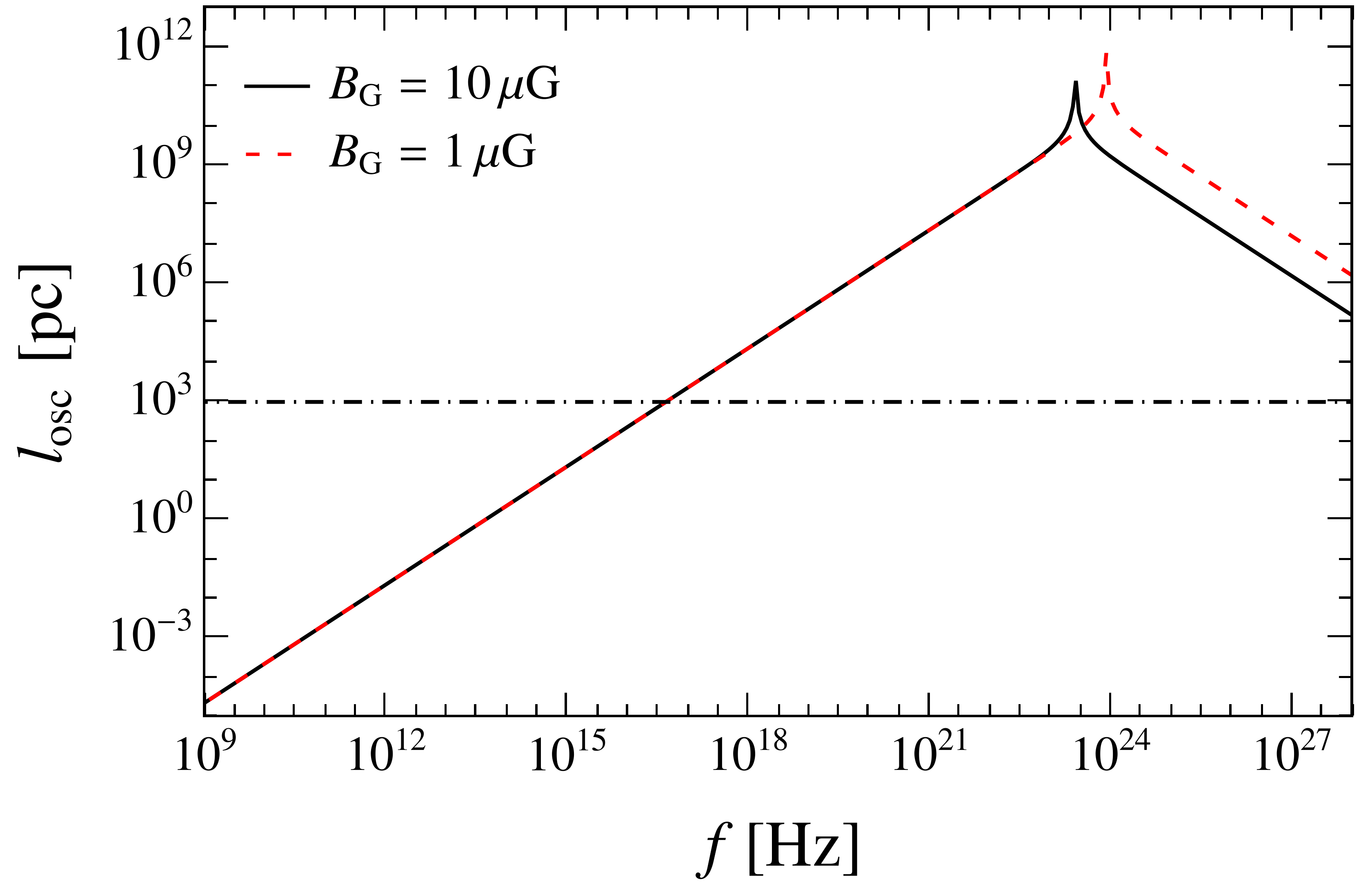}
    \caption{Graviton-photon oscillation length as defined in Sec.~\ref{subsec:HomogeneousField}. Here the electron number density is fixed at ${n_e=10^{-2}\,\cm^{-3}}$. The plot shows the behavior for two different values of the transverse Galactic magnetic field intensity.
    At $f\lesssim10^{23}\,\Hz$ the oscillation length is determined by $\Delta_{\rm pl}$, while at higher frequencies $\Delta_{\rm CMB}$ plays the dominant role~(see text for details).
    }
    \label{Fig:OscLength}
\end{figure}

The behavior of the oscillation length $l_{\rm osc}=2\pi/\Delta_{\rm osc}$ is displayed in Fig.~\ref{Fig:OscLength}. In this figure, we assumed an electron number density $n_e=10^{-2}\,\cm^{-3}$, which is compatible with the values expected in the Milky Way.  The low-frequency behavior is determined by $\Delta_{\rm pl}\sim\omega^{-1}$, so that the oscillation length increases linearly with the frequency. For frequencies $f\gtrsim10^{23}\,\Hz$ ($\omega\gtrsim4\,\GeV$) scattering over CMB photons starts to dominate over plasma effects and the oscillation length decreases as ${l_{\rm osc}\sim\Delta_{\rm CMB}^{-1}\sim\omega^{-1}}$. 

In the range of frequencies ${f\sim10^{18}-10^{28}\,\Hz}$, ${l_{\rm osc}\gg z}$ and many graviton-photon oscillations are possible on a typical Galactic scale $z\simeq20\,\kpc$. In this regime, graviton-photon mixing is strong and the conversion probability results to be energy independent:
\begin{equation}    
    \begin{split}
        P_{g\gamma}=(\Delta_{g\gamma}\,z)^2
        \simeq 7.8\times10^{-17}\,\left(\frac{B_{\rm T}}{1\,\mu \rm{G}}\right)^2\,\left(\frac{z}{10\,\kpc}\right)^2\,.
    \end{split}
\label{Eq:ProbHom}
\end{equation}

\section{Graviton-Photon conversions in Milky-Way magnetic fields}
\label{sec:MilkyWayConversions}
A realistic estimation of the conversion probability within the Milky Way depends on the morphology of the model assumed to describe the Galactic magnetic field. As a first approximation, we neglect the contribution from the turbulent component of the field, coherent over small scales $l_{\rm corr}\sim10-100\,{\rm pc}$~\cite{Iacobelli:2013fqa,Haverkorn:2008tb,10.1093/mnras/262.4.953,Carenza:2021alz}, since it is affected by large uncertainties (see Sec.~\ref{sec:RegVSTurb} for further details). In this regard, we highlight that the previous analyses dealing with graviton-photon conversions in the broadband CPB spectrum considered the turbulent component only~\cite{Ito:2023nkq}.

\begin{figure}[t!]
    \centering
    \includegraphics[width=1\columnwidth]{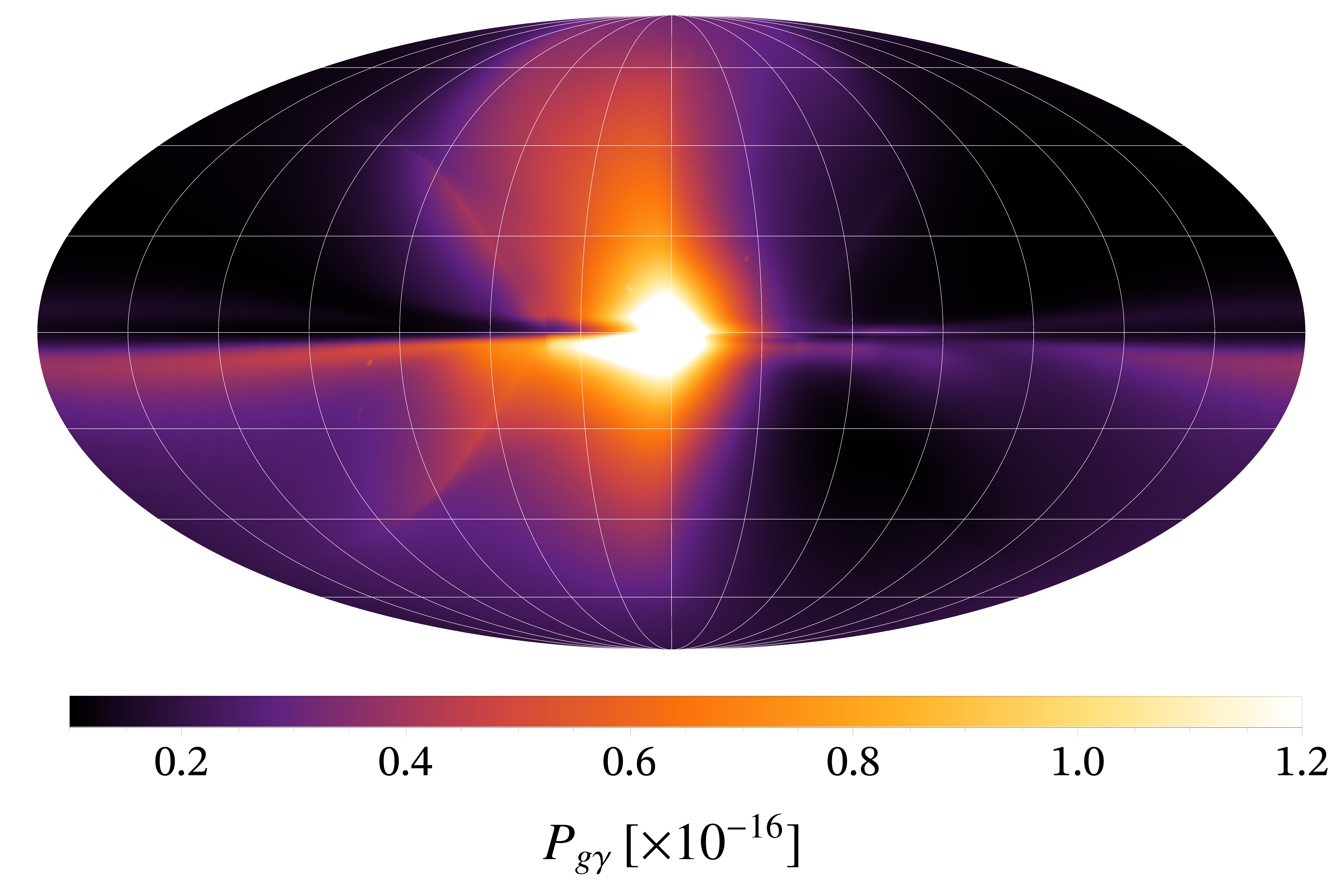}
    \caption{All-sky map for graviton-photon conversion probability in the strong mixing regime $f>10^{19}\,\Hz$ for the Jansonn-Farrar Milky-Way magnetic field model. The Figure is shown in Mollweide projection. Horizontal and vertical lines are equally spaced by $\Delta\theta=30^\circ$.}
    \label{Fig:SkyMap}
\end{figure}

In this work, we adopt the Jansson-Farrar model~\cite{Jansson:2012pc} as benchmark model for the Milky Way, regular magnetic field, which takes into account a disk field and an extended halo field with an out-of-plane component, based on WMAP7 Galactic synchrotron emission map~\cite{Gold:2010fm} and extra-galactic Faraday rotation measurements. In particular, we employ the updated parameters given in Table~C.2 of Ref.~\cite{Planck:2016gdp}
(“Jansson12c” ordered fields), matching the polarized synchrotron and dust emission measured by the Planck satellite~\cite{Planck:2015mrs,Planck:2015qep,Planck:2015zry}. Fig.~\ref{Fig:SkyMap} displays the all-sky map of the conversion probability obtained by integrating numerically the EoMs in Eq.~\eqref{Eq:EOM} in the strong mixing regime $\omega\sim100\,\MeV$ and by considering a beam propagation starting from $z=20\,\kpc$ from Earth. Since in this regime the conversion probability is energy independent, it strictly reproduces the morphology of the magnetic field model employed. In particular, it is significantly stronger around the Galactic center $\ell\in[-7.5^{\circ},7.5^{\circ}]$ and $b\in[-15^{\circ},15^{\circ}]$, where $\ell$ and $b$ are the Galactic longitude and latitude, respectively. In the range $\ell\in[-30^{\circ},30^{\circ}]$, out-of-plane components of the Jansson-Farrar model makes possible to have a large amount of power also at high latitudes. On the other hand, for Galactic longitudes $\ell\in[-180^{\circ},-30^{\circ}]\cup[30^{\circ},180^\circ]$, we observe a significant conversion probability just along the Galactic disk.

\begin{figure}[t!]
    \centering
    \includegraphics[width=1\columnwidth]{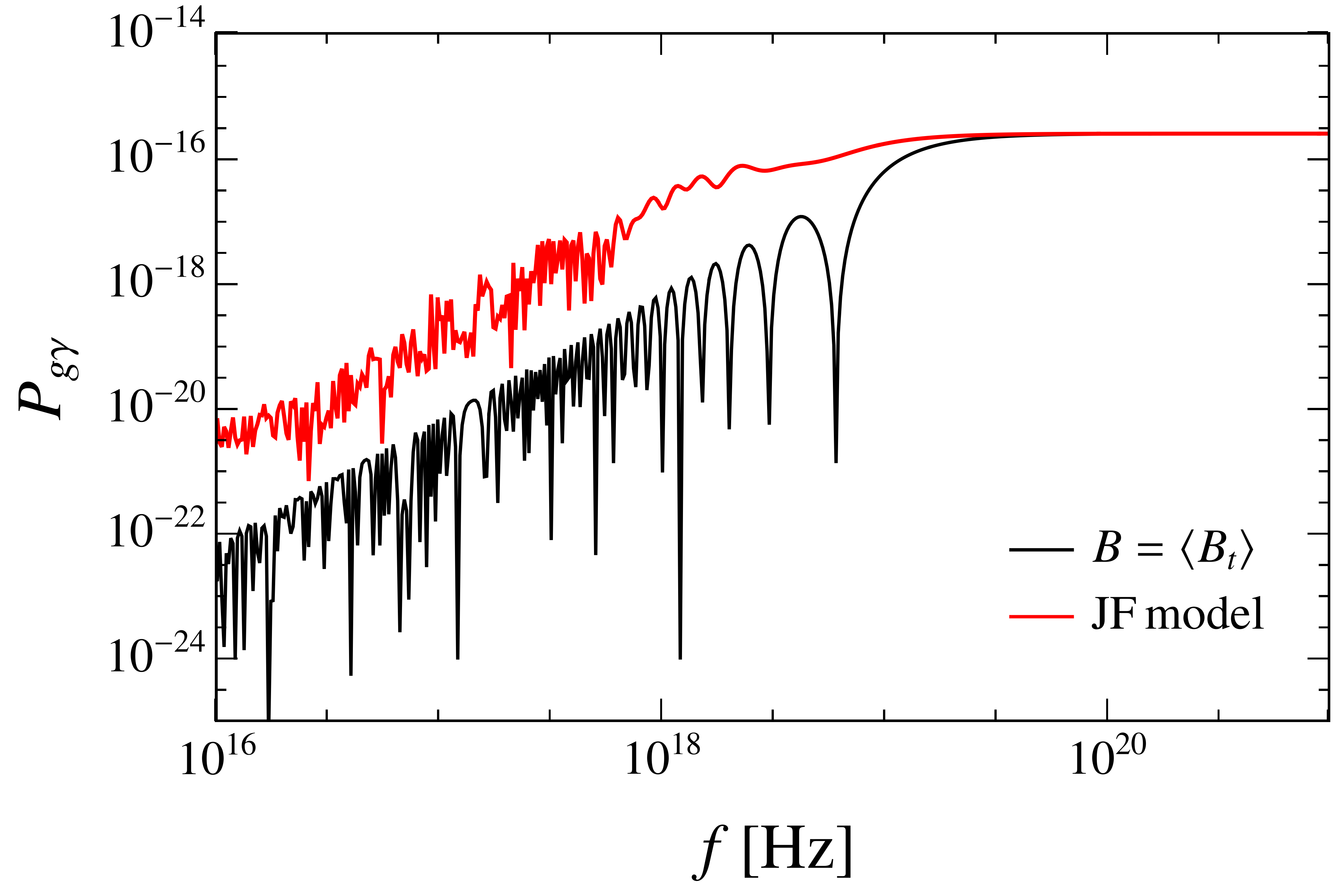}
    \caption{Graviton-photon conversion probabilities along the direction of the Galactic centre. In particular, the black line refers to a constant and homogeneous Galactic magnetic field ${\langle B_T\rangle=0.91\,\mu}$G and electron number density ${\langle n_e\rangle =7.1\times10^{-2}\,\cm^{-3}}$, while the red line depicts the probability computed by solving numerically Eq.~\eqref{Eq:EOM}, which takes into account the full morphology of the Jansson-Farrar~(JF) model.}
    \label{Fig:Comparison}
\end{figure}

In order to compare the behavior of the numerical solutions of the EoMs and the analytical result in Eq.~\eqref{Eq:ProbHom}, it is convenient to introduce the average absolute value of the transverse magnetic field component along the line of sight considered
\begin{equation}
    \langle B_T\rangle^2=\frac{1}{L^2}\left(\left|\int_0^L B_x(z,l,b)dz\right|^2+\left|\int_0^L B_y(z,l,b)dz\right|^2\right)\,,
\end{equation}
where $B_x$ and $B_y$ are the magnetic field components parallel to $A_x$ and $A_y$, respectively. In this expression, we set $L=20\,\kpc$. Analogously, the average value of the electron density along the chosen line of sight can be defined as
\begin{equation}
     \langle n_e\rangle=\int_0^Ldz\,n_e(z,l,b)\,.
\end{equation}
Fig.~\ref{Fig:Comparison} displays the energy behavior of the conversion probability for the Jansson-Farrar magnetic field model, comparing the numerical solution of the EoMs (red line) and analytical expression in Eq.~\eqref{Eq:EOMhom} obtained by employing the average values $\langle B_T\rangle$ and $\langle n_e\rangle$ (black line). In particular, the line of sight assumed in this plot is the Galactic center direction
$(\ell,b)=(0^\circ,0^\circ)$, along which $\langle B_T\rangle\simeq0.91\,\mu$G and $n_e=7.1\times10^{-2}\,\cm^{-3}$. We observe that at frequencies $f>10^{19}\,\Hz$ the two approaches provide the same results. Actually, in this range of frequencies the oscillation length $l_{\rm osc}$ is larger than the Galactic magnetic field scale ($l_{\rm osc}>20\,\kpc$), and graviton-photon oscillations are not affected by the magnetic field morphology along the line of sight, which varies over scales $l\sim1\,\kpc$. On the other hand, at lower frequencies the detailed magnetic field structure becomes important and the results provided by the two approaches are different. In particular, for frequencies $f<10^{19}\,\Hz$ the conversion probabilities associated to the exact solution are always larger than the homogeneous magnetic field approximation. Indeed, as shown in Ref.~\cite{Marsh:2021ajy} for the axion case, the probability to observe axion-photon oscillations is essentially proportional to the $B$-field power spectrum, which is enhanced by the presence of inhomogeneities. Hence, since inhomogeneities are washed out in the constant magnetic field approximation, the conversion probability results to be reduced.  Nevertheless, in both cases the probability scales as $P_{g\gamma}\sim\omega^2$ in the range of frequencies $f<10^{18}\,\Hz$, where plasma effects play the dominant role (see Section~\ref{subsec:HomogeneousField}). 

It is worthy to observe that at  frequencies ${f\lesssim10^{16}\,\Hz}$, where the oscillation length is smaller than the Galactic magnetic field scale $l_{\rm osc}\ll 20\,\kpc$, the numerical solutions of Eq.~\eqref{Eq:EOM} start to be affected by numerical artifacts, due to loss of precision in computation of highly-oscillatory solutions. Therefore, for our comparison with data related to the CPB, we will employ the analytical expression in Eq.~\eqref{Eq:ProbConstB}. This choice is justified by the previous comparison between the two approaches, which assures that the constant magnetic-field approximation provides the most conservative limits on a diffuse photon excess.

\subsection{Turbulent component}
\label{sec:RegVSTurb}
\begin{figure}[t!]
    \centering
    \includegraphics[width=1\columnwidth]{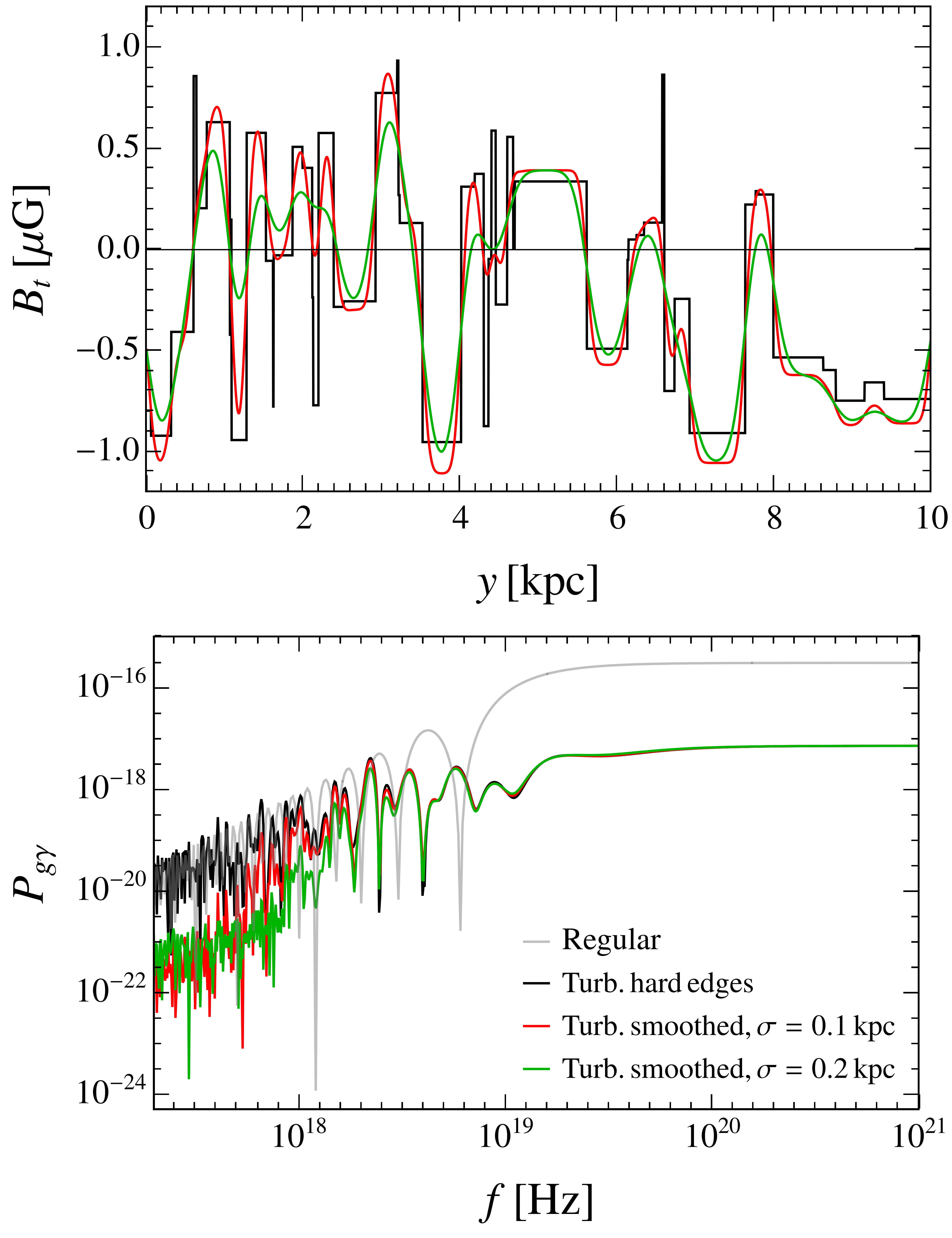}
    \caption{\textit{Upper Panel}: Representation of a possible configuration of the Galactic turbulent magnetic field component. Here we assume a cell model characterized by a maximum magnitude of the transverse component $|B_{\rm T}|=1\,\mu$G and the magnetic field is assumed to vary randomly over scales $l_{\rm corr}\sim 200\,$pc. In particular, the black lines refer to fields which change value suddenly moving from one cell to the other (hard edges), while the red and green lines have been obtained by smoothing the magnetic field variations at the boundary of each cell (see text for details). \textit{Lower Panel}: Graviton-photon conversion probabilities associated to the three turbulent magnetic field configurations displayed in the upper panel. For comparison, the regular component assuming an homogeneous field with $B_{\rm T}=1\,\mu$G is displayed as a a light-gray solid line. 
     }
    \label{Fig:HardSoft}
\end{figure}

Radio observations of synchrotron emission and polarization revealed the existence of a small-scale component of the Milky Way magnetic field related to the turbulent Inter-Stellar Medium (see, e.g., Ref.~\cite{Reich:2001fb}). This component is known as the random or turbulent magnetic field, and it is expected to have an amplitude comparable to that of the regular magnetic field, but a much shorter correlation length, $l_{\rm corr}\sim \mathcal{O}(10-100)\,$pc~\cite{Iacobelli:2013fqa,Haverkorn:2008tb,Malkov:2010yq}. In order to model a turbulent component of the magnetic field, the simplest possible choice is to adopt a \textit{cell model}, in which the magnetic field along a given line-of-sight is taken to be constant within a series of domains with (possibly) variable length and magnetic field intensity~\cite{Conlon:2015uwa,Marsh:2017yvc,Marsh:2021ajy}. Clearly, this toy model does not correspond to magnetic fields actually realized in nature, but it captures the most important aspects of graviton-photon oscillations in magnetic fields characterized by a short correlation length with respect to the size of the whole region considered $L\sim\mathcal{O}(10)\,\kpc$. As a first approximation, it is possible to assume that the intensity of the random magnetic field changes suddenly moving from one cell to the other so that the different magnetic domains appear to be separated by ``hard edges''. On the other hand, a more realistic choice for the field configuration would require model with smoothed edges between the domains. In the context of axion-photon conversions, the difference between this model and the cell model was investigated in Refs.~\cite{Galanti:2018nvl,Marsh:2021ajy}, following analytical arguments. As an illustrative example, the upper panel of Fig.~\ref{Fig:HardSoft} show a possible configuration of such a field, assuming a cell scale $l_{\rm corr}\sim 200\,$pc and transverse magnetic fields amplitudes which vary randomly between $B_{\rm T}\in[-1,1]\,\mu$G. The solid black line refer to a possible hard-edges configuration, while red and green solid lines depict two smoothed versions of the considered field, obtained as 
\begin{equation}
    B_{\rm smooth}(x)=\int dy\,\frac{1}{\sqrt{2\pi\sigma^{2}}}e^{-\frac{(x-y)^{2}}{2\sigma^{2}}}B_{\rm T}(y)\,,    
\end{equation}
where $B_{\rm T}$ is the cell model magnetic field transverse component and $\sigma$ is a smoothing scale. In particular, we assumed $\sigma=100\,$pc and $\sigma=200\,$pc for red and green lines, respectively.

Starting from Eq.~\eqref{Eq:EOMhom}, it is possible to calculate, at the lowest order in $1/M_{P}$, the transition amplitude for graviton-photon conversion as~\cite{Marsh:2021ajy}
\begin{equation}
    \begin{split}
\mathcal{A}^{(i)}_{g\to\gamma}&\simeq \int_{z_{0}}^{z}dz'\, \Delta_{g\gamma,i}\,e^{-i\int_{z_{0}}^{z'} dz''\Delta_{i}}=\\
&=\int d\Phi_{i}\, \frac{\Delta_{g\gamma,i}}{|d\Phi_{i}/dz'|}e^{-i\lambda \Phi_{i}}
\end{split}
\end{equation}
where $i=x,y$ is associated to a given $\Delta_{g\gamma,x}=\Delta_{g\gamma}s_{\phi}$ or $\Delta_{g\gamma,y}=\Delta_{g\gamma}c_{\phi}$. Here we defined $\lambda\Phi_{i}(z')=\int_{0}^{z'}dz''\,\Delta_{i}$ and $\lambda=1/f$. Then, following the discussion in Ref.~\cite{Marsh:2021ajy}, one realizes that the conversion probability is proportional to the power spectrum of $B_{\rm T}/|d\Phi_{i}/dz'|$ on scales dictated by $1/f$, which encodes the properties of magnetic field and plasma and their correlation lengths. The lower panel of Fig.~\ref{Fig:HardSoft} shows the behavior of the conversion probability for the hard edges and smoothed configurations considered. We can appreciate that at frequencies $f>10^{19}\,\Hz$ the graviton-photon oscillation length is much larger than the domain size $l_{\rm osc}\gg 200\,$pc. Hence, the oscillation probability is not sensitive to the detailed structure of the turbulent magnetic field and for all the three models the probability can be simply estimated as~\cite{Carenza:2021alz}
\begin{equation}
    P_{\rm turb}\simeq N\,\Delta P=\langle B_{\rm T}\rangle^2\,L\,l_{\rm corr}\,
\end{equation}
where $N=L/l_{\rm corr}$ is the number of cells and $\Delta P$ is the conversion probability in a single domain. Thus, in this frequency range, one expects the turbulent magnetic field contribution to be suppressed by a factor $~l_{\rm corr}/L$ with respect to the regular one. This behavior can be appreciated by comparing the probabilities for turbulent configurations to the results obtained for the regular component associated to a homogeneous field with $B_{\rm T}=1\,\mu$G, displayed in Fig.~\ref{Fig:HardSoft} as a light-gray solid line.
At lower frequencies the oscillation length is comparable or smaller than turbulent field scales and the conversion probability is strongly affected by the magnetic field structure over small scales. In particular, hard edges at cell boundaries enhance the magnetic field power spectrum leading to higher conversion probabilities with respect to the smooth-edges configurations. Thus, in agreement with the results of Refs.~\cite{Kartavtsev:2016doq,Ito:2023nkq}, the hard-edges assumption leads to an overestimation of the conversion probability by a factor $\sim N$ in the range of frequencies where $P_{g\gamma}\sim\omega^{2}$. In this regime, the turbulent component with hard-edges configuration could provide a contribution comparable to the regular one, while, in case of the smoothed configurations, this contribution can be safely neglected. Thus, the impact of the turbulent component in the oscillatory regime results to be strongly dependent on the assumptions made on the $B$-field power spectrum.

In general, the previous discussion suggests that the turbulent magnetic field contribution might give a sizeable contribution at low frequencies, where small-scale anisotropies could power the field power spectrum. Indeed, we showed that the detailed structure of the turbulent component is relevant when we approach to short oscillation lengths. However, the small-scale structure of the turbulent $B$-field, which is essential to determine the associated power spectrum, is still largely unclear. Therefore, in order to make our results less arbitrary,
in the following we will focus just on the regular Milky-Way magnetic field, neglecting the contribution from the turbulent one. We stress that this is quite orthogonal to the treatment of~\cite{Ito:2023nkq}.

\section{Distortions in the CPB spectrum}
\label{sec:Distortion}
A diffuse and isotropic GWB at frequencies $f>10^{9}\,\Hz$ can be converted into photons inside the Galactic magnetic field, giving rise to a diffuse photon flux. In particular, a GWB is associated to the flux of gravitons per unit frequency and solid angle which can be expressed as~\cite{Maggiore:2000gv,Dandoy:2024oqg} 
\begin{equation}
    \begin{split}
        \frac{d^2F_g}{dfd\Omega}&= \frac{1}{4\pi}\frac{d\rho_{\rm GW}}{df}\\
        &=\frac{\pi}{4}\,M_{\rm pl}^2\,f\,h_c^2\,,
    \end{split}
\label{Eq:GravitonFlux}
\end{equation}
where we have introduced the GW spectral energy density $d\rho_{\rm GW}/df$ and the characteristic strain $h_c$~\cite{Maggiore:2000gv}, and $\Omega$ is the solid angle. Then, the induced photon flux can be evaluated as
\begin{equation}
    \frac{d^2 F_{g \rightarrow \gamma}}{df d\Omega}(f,\omega)=\frac{d^2F_g}{dfd\Omega}(f)\,P_{g\gamma}(f,\Omega)\,.
\label{Eq:PhotonFlux}
\end{equation}
In particular, Eq.~\eqref{Eq:PhotonFlux} points out that the spatial distribution of the induced photon flux reproduces the morphology of the Milky-Way magnetic field, encoded in the conversion probability, $P_{g\gamma}$.\\
Eq.~\eqref{Eq:PhotonFlux} implies that a too large GWB might give rise to a relevant diffuse photon flux, imprinting distortions in the observed CPB. Hence, current observations of the CPB naturally imply limits on the considered GWB. In the next Section, we briefly review the set of CPB data employed in this work to derive observational constraints on the GWB.

\subsection{Observations and emission models}
\label{subsec:CPB}
The set of the CPB measurements employed in this work is taken from Ref.~\cite{Hill:2018trh}, which gathered the most recent measurements of the sky-averaged intensity spectrum from radio to $\gamma$-ray energies, reviewing the current understanding of the CPB. In the following we briefly describe the datasets considered in the different frequency ranges. Moreover, current observations of the CPB are well explained by different, astrophyisical emission models in most frequency bands. Effectively, accounting for the astrophysical contributions to the CPB reduces the room left for a graviton-photon conversion signal, as we will show in Sec.~\ref{subsec:Results}.
For each section of the CPB we report some illustrative models aiming to explain the observed emission in the frequency band considered, which will be used later on.

\begin{itemize}
    \item \textit{Cosmic radio background (CRB), $[10^9-10^{10}\,\Hz]$.} Data in this frequency band come from a combination of ground and balloon-borne instruments. Data reported in Fig.~\ref{Fig:CPB} include measurements of the ARCADE experiment~\cite{Fixsen:2004hp} and its successor, ARCADE-2~\cite{Fixsen:2009xn}, the TRIS experiment~\cite{Zannoni:2008xx}, and VLA~\cite{Condon:2012ug,Vernstrom:2013vva} and ATCA~\cite{Vernstrom:2014uda} telescopes. These data are well-interpolated by the low-frequency tail of a black-body spectrum at the CMB temperature $T_{\rm CMB}=2.7255\pm0.0005$ K~\cite{Fixsen:2009ug}.

    \item \textit{Cosmic microwave background (CMB), ${[10^{10}-10^{12}\,\Hz]}$.} High-precision CMB spectral measurements have been provided by the FIRAS instrument over the COBE satellite~\cite{Fixsen:1996nj}. This spectrum is described as a blackbody radiation left over from the hot and early phase of the Universe, characterized today by a temperature $T=T_{\rm CMB}$.

    \item \textit{Cosmic infrared background (CIB), ${[10^{12}-10^{14}\,\Hz]}$.} The CIB radiation is expected to be mainly emitted by dust heated by stars within unresolved galaxies.
    Since objects in our Solar System significantly contribute to radiation emission in this band, this region of the spectrum results to be particularly difficult to measure and experimental observations mostly provide upper and lower limits. Measurements in this frequency band comes from PACS~\cite{Berta:2011xi}, SPIRE~\cite{Bethermin:2012jd}, BLAST~\cite{Marsden:2009qs}, SCUBA-2~\cite{Chen:2013gwa}, DIRBE~\cite{Odegard:2007qy}, FIS~\cite{Matsuura:2010rb} and PLANCK~\cite{Planck:2013wqd}. Together with these measurements, Ref.~\cite{Hill:2018trh} provides a vast catalog of upper and lower limits obtained by summing up the contribution of resolved sources. An attempt to model CIB and the cosmic optical background (COB) is performed in Ref.~\cite{Dominguez:2010bv}, by making a weighted sum of EBL emission inferred from spectra of observed galaxies.

\begin{figure}[t!]
    \centering
    \includegraphics[width=1\columnwidth]{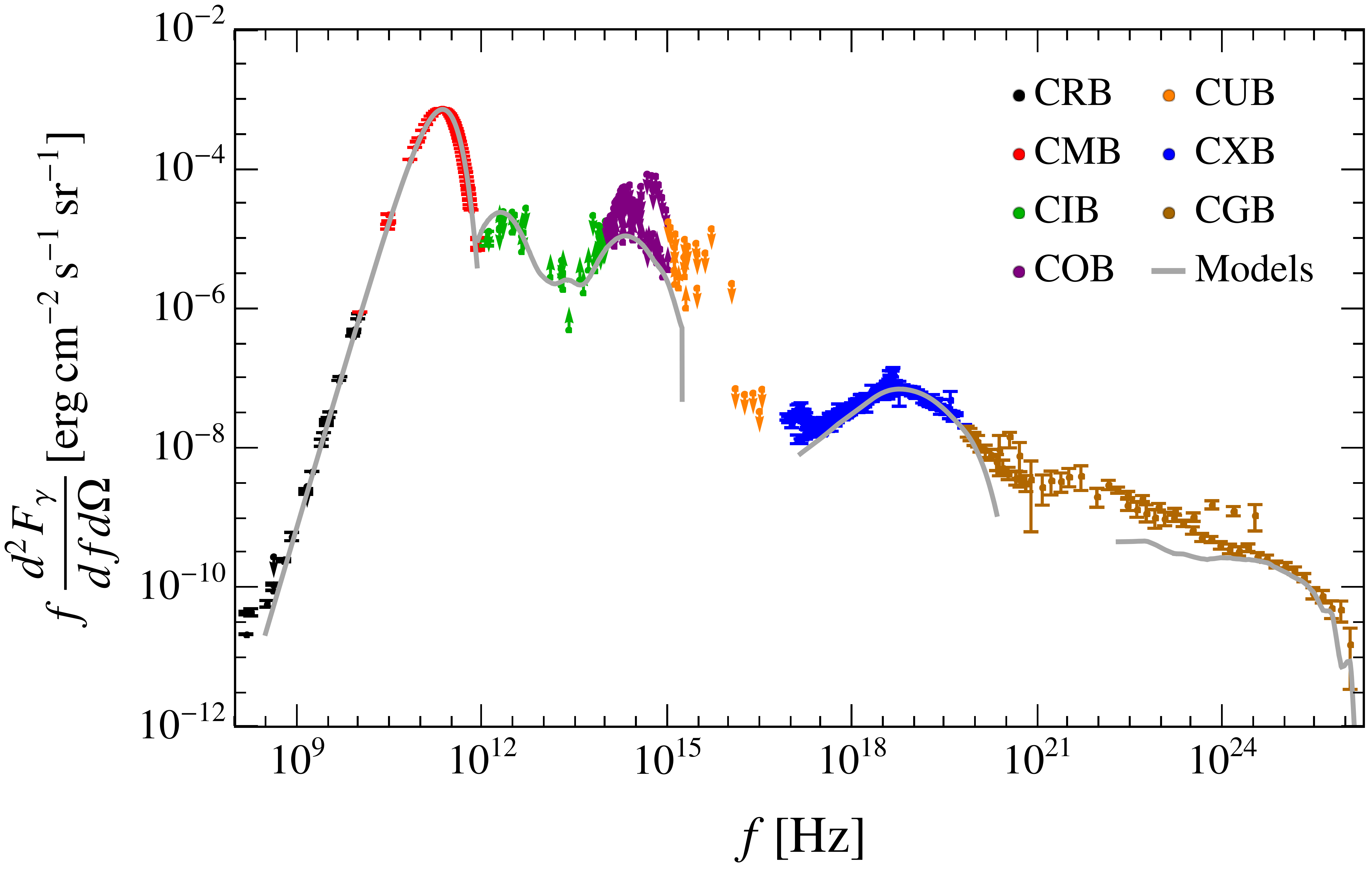}
    \caption{Full spectrum of the CPB in the range of frequencies considered in this work. All the data employed in this figure are taken from Ref.~\cite{Hill:2018trh}. The different colors refer to the different regions of the spectrum described in Sec.~\ref{subsec:CPB}. In particular, upward and downward arrows display upper and lower limits on the CPB intensity. We also show in gray the emission models reported in Sec~\ref{subsec:CPB}, introduced to explain observations in the different ranges of frequencies.}
    \label{Fig:CPB}
\end{figure}

    \item \textit{Cosmic optical background (COB), ${[10^{14}-10^{15}\,\Hz]}$.} The COB spectrum is dominated by the emission coming directly from stars. Also in this case, Milky-Way dust emission generates an unavoidable foreground for optical-frequency measurements and direct estimations of the COB are particularly challenging. Hence, in this energy range, current direct observations just place upper and lower limits on the photon intensity, as reported in Ref.~\cite{Hill:2018trh}. As for the CIB spectrum, we characterize the emission in optical frequencies via the model in Ref.~\cite{Dominguez:2010bv}.

    \item \textit{Cosmic ultraviolet background (CUB), ${[10^{15}-10^{16}\,\Hz]}$.} The origin of this background can be linked to hot young stars and interstellar nebulae. Actually, the CUB  spectrum is poorly studied because UV light is efficiently absorbed by neutral hydrogen and Earth's atmosphere. Since data in these frequency range contain systematic uncertainties hard to quantify, observations provide only upper and lower limits~\cite{Hill:2018trh}.

     \item \textit{Cosmic X-ray background (CXB), ${[10^{16}-10^{19}\,\Hz]}$.} Diffuse emission in the X-ray band is attributed to emission of bremsstrahlung photons in the hot accretion disks around active galactic nuclei~(AGN)~\cite{Comastri:1994bz}. Moving from soft to hard X-ray spectra, datasets characterizing these wavelengths have been provided by ROSAT and ASCA experiments~\cite{Miyaji:1998ez}, the SWIFT X-ray telescope~\cite{Moretti:2008hs}, XMM-Newton~\cite{DeLuca:2003eu}, CHANDRA~\cite{Cappelluti:2017miu}, RXTE~\cite{Revnivtsev:2003wm}, BeppoSAX~\cite{Frontera:2006zg}, INTEGRAL~\cite{Turler:2010pm}, BAT telescope~\cite{Ajello:2008xb} and HEAO~\cite{Gruber:1999yr}. We employ the model in Ref.~\cite{Ueda:2014tma} as benchmark AGN emission model in this frequency range.

     \item \textit{Cosmic $\gamma$-ray background (CGB), ${[10^{20}-10^{26}\,\Hz]}$.} The CGB covers all the range of frequencies ${f\gtrsim10^{19}\,\Hz}$. At these energies, the dominant emission is expected to come from quasars and blazars~\cite{Ajello:2015mfa}, supernova explosions~\cite{Ruiz-Lapuente:2000omh} and star-forming galaxies~\cite{Tamborra:2014xia}. CPB data employed at these wavelengths comes from observations of the GRS experiment~\cite{doi:10.1063/1.53933},  COMPTEL~\cite{Weidenspointner:2000aq}, EGRET~\cite{Strong:2004ry}, and \textit{Fermi}-LAT~\cite{Fermi-LAT:2014ryh}. In order to avoid to saturate photon intensity, we assume blazar emission as modelled in Ref.~\cite{Fornasa:2015qua} as only emission source in this frequency band.
\end{itemize}

At higher frequencies $f\gtrsim10^{28}\,\Hz$, very-high energies photons undergo pair production absorptions by background low energy photons from EBL ${\gamma_{\rm VHE}+\gamma_{\rm EBL}\rightarrow e^++e^-}$. Thus, absorption effects are not negligible anymore and graviton-photon propagation has to be described by means of a Liouville equation~(see Ref.~\cite{Mastrototaro:2022kpt} for a recent work considering the axion case). Therefore, in this work, we decided to neglect measurements of the photon background by the Large High Altitude Air Shower Observatory (LHAASO)~\cite{LHAASO:2023gne}.

\begin{figure*}[t!]
    \centering
    \includegraphics[scale=0.65]{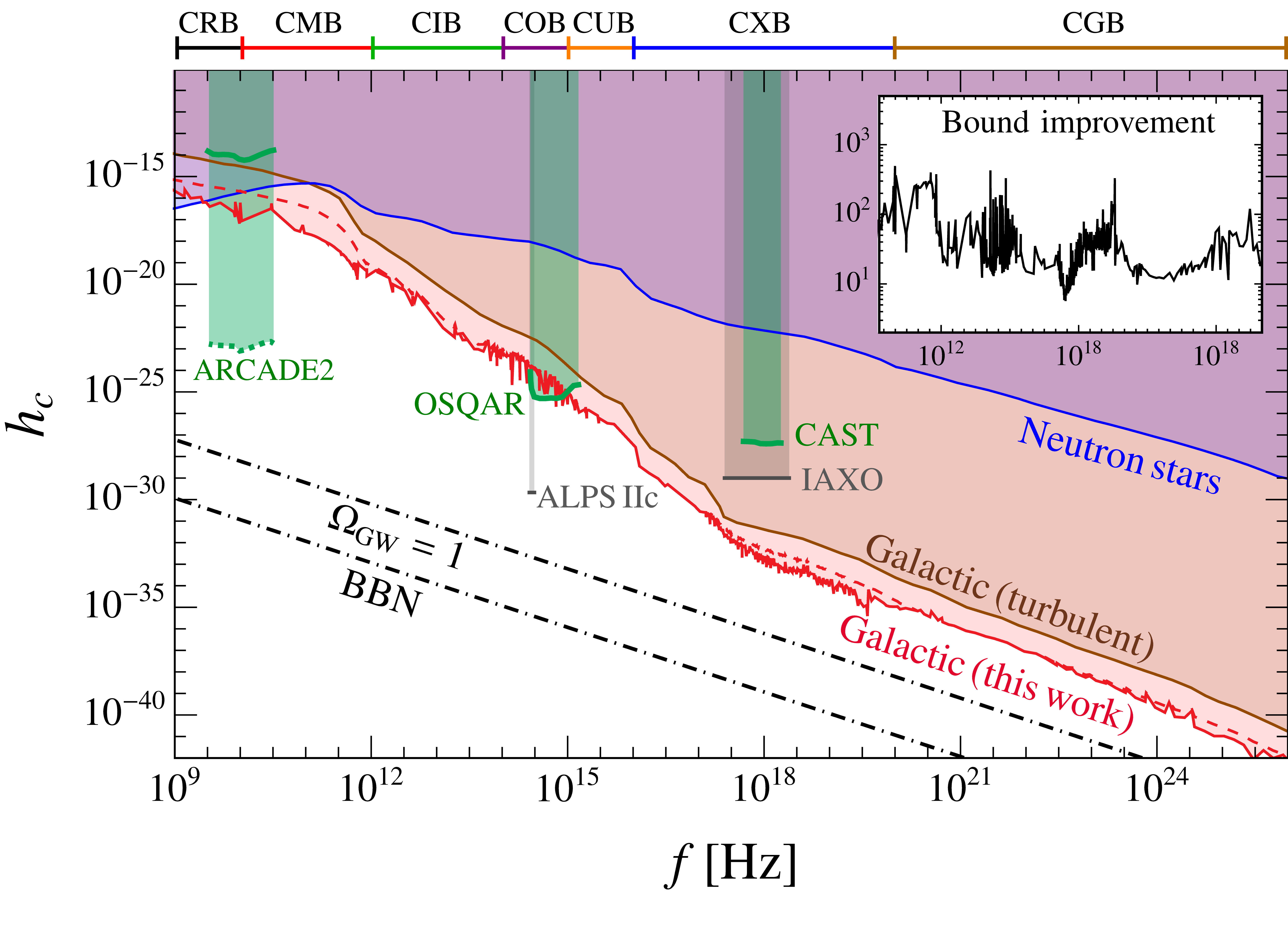}
    \caption{Summary plot of current bounds on the characteristic strain associated to a GWB at frequencies $f\in[10^{9},10^{26}]\,\Hz$. The upper part of the frame shows the regions of the CPB spectrum associated to each frequency range. The exclusion limits introduced in this work are displayed as solid and dashed red lines for the fiducial and the most conservative bound, respectively. We report limits introduced in Ref.~\cite{Ito:2023nkq} by employing a turbulent Galactic magnetic field component characterized by and average magnitude $B_{\rm T}=1\,\mu$G and a correlation length $l_{\rm corr}\sim100\,$pc~(brown solid line). In particular, the subplot inset shows the factor of improvement between the bound introduced in this work and the constraint of Ref.~\cite{Ito:2023nkq} in the whole range of frequencies considered. In addition, we also display limits from Ref.~\cite{Dandoy:2024oqg} by considering conversions in the Galactic NS population~(blue solid line). Green areas refer to regions excluded by the OSQAR and CAST experiments~\cite{Ejlli:2019bqj}, as well as the exclusion region set in Ref.~\cite{Domcke:2020yzq} by employing the ARCADE2 telescope. Here, solid and dashed green lines refer to the most conservative and the most aggressive choice of parameters for primordial magnetic fields. Moreover, we report in gray the projected sensitivities of the ALPS IIc and IAXO future experiments~\cite{Ejlli:2019bqj}. Finally, dot-dashed black lines represent the BBN bound and the limit obtained by requiring that the GW energy density does not saturate the total energy density of the Universe~(see Sec.~\ref{subsec:Results} for further details).}
    \label{Fig:SummaryPlot}
\end{figure*}

\subsection{Constraints on the GWB}
\label{subsec:Results}

Conversions of a large GWB may have given rise to an observable diffuse photon flux in the range of frequencies considered $10^{9}\,\Hz\lesssim f\lesssim10^{26}\,\Hz$.
Nevertheless, as discussed in the previous Section, observations of the CPB are well explained by the emission models in the different energy ranges. Thus, if conversions of GWs  contribute to a diffuse photon background, the flux intensity associated to this contribution has to be smaller than the relative discrepancies between measurements and the emission models explaining observations in a given frequency range. Namely, we require
\begin{equation}
    \frac{1}{4\pi}\frac{dF_{g\rightarrow \gamma}}{df}\Bigg{|}_{f=f_i}\lesssim\left|\frac{d^2F_{\gamma,i}^{\rm exp}}{dfd\Omega}-\frac{d^2F_{\gamma,i}^{\rm th}}{dfd\Omega}\right|
\label{Eq:BoundCondition}
\end{equation}
where ${d^2F_{\gamma,i}^{\rm exp}}/{dfd\Omega}$ and ${d^2F_{\gamma,i}^{\rm th}}/{dfd\Omega}$ are the measured and theoretically predicted CPB at frequency $f_i$, and ${dF_{g \rightarrow\gamma}}/{4\pi df}$ is the photon flux induced by graviton-photon conversions averaged over the whole solid angle. 
The theoretical predicted CPB contributions are the ones detailed in the previous Section, corresponding to the astrophysical models in Fig.~\ref{Fig:CPB}. By computing the photon flux induced by conversions by means of Eq.~\eqref{Eq:PhotonFlux}, and from the residuals computed in Eq.~\ref{Eq:BoundCondition}, we can derive upper limits on the GW strain.

Fig.~\ref{Fig:SummaryPlot} displays our new constraints on the characteristic strain associated to a GWB at high frequencies $f\gtrsim10^{9}\,\Hz$. In particular, the fiducial constraint derived in this work is displayed as solid red line. 
A more conservative limit on the characteristic strain can be obtained without assuming any model for the emission of the radiation contributing to the CPB. In this case, constraints on the characteristic strain can be derived by requiring that GW conversions do not produce a photon background larger than the observed fluxes 
\begin{equation}
    \frac{1}{4\pi}\frac{dF_{g \rightarrow \gamma}}{df}\Bigg{|}_{f=f_i}\lesssim\frac{d^2F_{\gamma,i}^{\rm exp}}{dfd\Omega}\,.
\end{equation}
In Fig.~\ref{Fig:SummaryPlot}, this conservative constraint is displayed as a dashed red line. We can appreciate that, in the conservative estimation, the bound relaxes of less than one order of magnitude in the CMB region, where the black body spectrum at $T_{\rm CMB}$ interpolates the data with a high precision level. On the other hand, the fiducial and the conservative limits are strictly comparable in the rest of the CPB spectrum, where data-to-model discrepancies are of the same order of magnitude of measurements.


\section{Discussion and Conclusions}
\label{sec:Discussion}

In this work we have studied the phenomenology associated to conversions of high frequency GWs into photons inside Galactic magnetic fields. For this purpose, we focused on the regular magnetic field component which has been modeled through the realistic Jansson-Farrar model, analyzing as well the impact of a possible turbulent component which is expected to be small. 

The comparison of our results to observations of the CPB in the range of frequencies ${10^9\lesssim f\lesssim10^{26}\,\Hz}$ allowed us to derive the strongest astrophysical bounds on the characteristic strain related to a GWB in this frequency range. 
Indeed, we reinforced at least by one order of magnitude the limits introduced in Ref.~\cite{Ito:2023nkq} (brown solid line), which considered the turbulent magnetic field component only. In this regard, the subplot in Fig.~\ref{Fig:SummaryPlot} shows the improvement factor between our bound and the constraint from Ref.~\cite{Ito:2023nkq} in the whole frequency range considered. In particular, our results are compared to the contours shown in Fig.~3 of Ref.~\cite{Ito:2023nkq} for the case of a turbulent field of average magnitude $B_{\rm T}$ and correlation length $l_{\rm corr}=100\,$pc.
This difference is manly due to the magnitude of conversion probabilities associated to magnetic fields coherent over Galactic scales, which result to be $\sim 2$ orders of magnitude larger than the case of turbulent fields with short correlation lengths~(see Sec.~\ref{sec:RegVSTurb}).
Furthermore, in the range $10^{9}<f<10^{12}\,\Hz$ our bounds are comparable and even stronger than the ones introduced in Ref.~\cite{Dandoy:2024oqg} by considering GW conversions inside magnetic fields of a Galactic neutron star population (solid blue line). For the sake of clarity, we remark that our results are obtained under the assumption of a one-dimensional propagation of the graviton-photon beam. While this can be a good first approximation for the phenomenology treated in this work, it might miss certain effects which emerge in a pure three-dimensional treatment (see, e.g., Ref.~\cite{McDonald:2024nxj} in which the authors study resonant graviton-photon conversion in strongly magnetized plasmas in neutron star magnetospheres). These interesting effects deserve a dedicated investigation that we plan for a future work.

Further developments for the phenomenology discussed in this work deal with the implementation of the most recent models of the large scale coherent magnetic field of the Milky Way, such as those proposed in Ref.~\cite{Unger:2023lob}.

In parallel to astrophysical arguments, many experimental ideas have been proposed to investigate graviton-to-photon conversions in the high-frequency range.
In Fig.~\ref{Fig:SummaryPlot} we report in green current experimental constraints on the characteristic strain from detectors designed for axion searches, such as OSQAR and CAST~\cite{Ejlli:2019bqj}, as well as the bound placed by employing ARCADE2 observations in searches for a GWB background formed in the period between reionization and recombination~\cite{Domcke:2020yzq}. We observe that constraints placed in this work are comparable to OSQAR limit at $2.7\times10^{14}\,\Hz\lesssim f\lesssim1.4\times10^{15}\,\Hz$ excluding $h_c\gtrsim6\times10^{-26}$, while they result to be five orders of magnitude stronger than limits from CAST experiment at $5\times10^{18}\,\Hz\lesssim f\lesssim1.2\times10^{19}\,\Hz$. In addition, we point out that the constraints in Ref.~\cite{Ejlli:2019bqj} might have been overestimated since the angular distribution of the signal has not been properly taken into account~(see the Supplemental Material of Ref.~\cite{Liu:2023mll} for further details). On the other hand, the ARCADE2 constraint introduced in Ref.~\cite{Domcke:2020yzq} strongly depends on the choice of the parameters describing primordial magnetic fields, which are affected by large uncertainties. Correspondingly, the excluded values of the strain vary from $h_c\gtrsim 10^{-14}$ (solid line) to $h_c\gtrsim 10^{-25}$ (dashed line). We also report in gray the projected sensitivities of the ALPS~IIc and IAXO experiments, which may be able to probe characteristic strains as low as $h_c\sim10^{-30}$ at $f\sim10^{14}\,\Hz$ and $h_c\sim10^{-29}$ at $f\sim10^{17}-10^{18}\,\Hz$, respectively~\cite{Ejlli:2019bqj}.

As a reference, we also report the Big-Bang nucleosynthesis (BBN) bound (black solid line), placed by requiring that the energy density stored in GWs $\rho_{\rm GW}$ does not exceed the total energy density associated to the effective number of massless degrees of freedom $N_{\rm eff}$~\cite{Pagano:2015hma}
\begin{equation}
    \rho_{\rm GW}\lesssim\frac{7}{8}\left(\frac{4}{11}\right)^{{4}/{3}}\,\Delta N_{\rm eff}\,\rho_{\gamma},
\end{equation}
where $\rho_{\gamma}$ is the current photon energy density and $\Delta N_{\rm eff}\lesssim0.3$ from CMB and BBN experiments~\cite{Planck:2018vyg,Cyburt:2015mya}. It is apparent that the BBN constraint is many orders of magnitude stronger than astrophysical bounds, especially in the range $f<10^{18}\,\Hz$. However, it is worthy to highlight that this bound does not apply to GWs sourced after the CMB decoupling, such as the scenario considered in~\cite{Domcke:2020yzq}. 

On the other hand, an unavoidable constraint to consider is the limit obtained by requiring that GWs do not saturate the total energy budget of the Universe, i.e. $\rho_{\rm GW}<\rho_{\rm c}$. This argument is again much more constraining than current experimental and astrophysical bounds in the range of frequencies $f<10^{18}\,\Hz$. Nevertheless, for higher frequencies, current sensitivity of telescopes and experiments searching for cosmic electromagnetic radiation are not far away from cosmological bounds. Hence, future developments in X- and $\gamma$-ray astronomy could open intriguing perspectives for the detection of a very-high-frequency GWB. 
In particular, at X-ray energies the future, ESA approved, {\it Athena} X-ray observatory is expected to push down the 
threshold for point-source detection~\cite{Cucchetti:2018izp}. This will imply an updated, more constraining measurements of the remaining 
diffuse signals as well as it will enable a better understanding of the population
contributing to this diffuse emission, boosting the sensitivity to exotic signatures. 
Similar arguments hold true for future MeV missions, like the programmed COSI telescope.\footnote{\url{https://cosi.ssl.berkeley.edu/instrument/science/}}
Finally, while at GeV -- TeV energies the future High Energy cosmic-Radiation Detection (HERD) facility\footnote{\url{https://herd.ihep.ac.cn/index.shtml}}, and the Cherenkov Telescope  Array (CTA) with its planned extragalactic 
survey~\cite{CTAConsortium:2017dvg} will shed further light on the composition of the extragalactic $\gamma$-ray background.

\section*{Acknowledgments}
We thank Pasquale Serpico for useful discussion on the topic, and Riley Hill for providing us the full dataset of CPB measurements. AL kindly thanks LAPTh for their hospitality during the development of this project. This article is based upon work from COST Action COSMIC WISPers CA21106, supported by COST (European Cooperation in Science and Technology).
The work of AM and AL was partially supported by the research grant number 2022E2J4RK ``PANTHEON: Perspectives in Astroparticle and
Neutrino THEory with Old and New messengers" under the program PRIN 2022 funded by the Italian Ministero dell’Universit\`a e della Ricerca (MUR). 
This work is (partially) supported
by ICSC – Centro Nazionale di Ricerca in High Performance Computing.
The work of PC is supported by the European Research Council under Grant No.~742104 and by the Swedish Research Council (VR) under grants  2018-03641 and 2019-02337.

\bibliographystyle{bibi.bst}
\bibliography{references.bib}

\end{document}